\documentclass[a4paper,11pt,preprintnumbers]{article}
\pdfoutput=1 

\usepackage{jcappub} 

\usepackage{booktabs}
\usepackage[T1]{fontenc} 
\usepackage{color}
\usepackage{graphicx, multirow, soul, url, amsmath, amsfonts, amssymb, mathrsfs}
\usepackage[all]{hypcap}
\usepackage{bbm}
\usepackage{cancel}
\usepackage{braket}
\usepackage[normalem]{ulem}
\usepackage{slashed}
\usepackage[dvipsnames]{xcolor}
\usepackage{multicol, blindtext}
\usepackage[section]{placeins}
\usepackage[version=3]{mhchem}
\usepackage{caption}
\usepackage{lipsum}
\usepackage{hyperref}
\usepackage{float}
\usepackage{mathtools}
\usepackage{footnote}
\usepackage{overpic, bm, epsfig}
\usepackage[splitrule]{footmisc}
\usepackage{bbold}
\usepackage{amsbsy}
\usepackage[export]{adjustbox}
\usepackage{units}
\usepackage{commath}
\usepackage{subcaption}

\newcommand{\be}{\begin{equation}}
\newcommand{\ee}{\end{equation}}
\newcommand{\bea}{\begin{eqnarray}}
\newcommand{\eea}{\end{eqnarray}}
\newcommand{\ba}{\begin{aligned}}
\newcommand{\ea}{\end{aligned}}
\newcommand{\fBH}{f_\mathrm{PBH}}

\newcommand{\equaref}[1]{Eq.~(\ref{#1})}
\newcommand{\equasref}[2]{Eqs.~(\ref{#1})~and~(\ref{#2})}

\newcommand{\figsref}[2]{Figures.~\ref{#1}~and~\ref{#2}}

\newcommand{\IPPP}{Institute for Particle Physics Phenomenology, Durham University, South Road, DH1 3LE, Durham, United Kingdom}
\newcommand{\AMS}{Department of Physics, University of Massachusetts, Amherst, MA 01003, USA}
\newcommand{\IFT}{Departamento de Física Teórica and Instituto de Física Teórica UAM/CSIC,
Universidad Autónoma de Madrid, Cantoblanco, 28049 Madrid, Spain}

\title{Baryogenesis via Asymmetric Evaporation of Primordial Black Holes}

\author[a]{Joaquim Iguaz Juan\,,}
\author[b]{Yuber F.~Perez-Gonzalez\,}
\author[c]{and Jessica Turner}

\affiliation[a]{\AMS}
\affiliation[b]{\IFT}
\affiliation[c]{\IPPP}

\emailAdd{jiguazjuan@umass.edu}
\emailAdd{yuber.perez@uam.es}
\emailAdd{jessica.turner@durham.ac.uk}

\preprintnumber{\newline IFT-UAM/CSIC-25-84, IPPP/25/51}

\abstract{
    We revisit baryogenesis from the asymmetric evaporation of light primordial black holes, focusing on scenarios where gravitational effects induce a matter antimatter asymmetry. In particular, we consider a higher-dimension operator coupling the Kretschmann scalar to a baryon-number-violating current which generates an effective chemical potential at the black hole horizon and leads to asymmetric Hawking radiation. Relative to earlier studies, we account for entropy dilution from evaporation, incorporate chemical potential dependent greybody factors and numerically track the fully coupled evolution of a PBH population in an expanding universe. We show that the observed baryon asymmetry can be reproduced within a viable region of parameter space for several PBH mass spectra including log‑normal, critical‑collapse, and power‑law distributions.
}


\begin{document}
\maketitle

\flushbottom

\def\thefootnote{\arabic{footnote}}
\setcounter{footnote}{0}

\section{Introduction}
The universe contains more matter than antimatter, with a baryon-to-entropy ratio,
$n_{\mathrm{B}} / s \simeq 9 \times 10^{-11}$
 \cite{Planck:2018vyg}. Explaining this asymmetry remains an open problem in fundamental physics and many mechanisms have been proposed (see Refs.~\cite{Riotto:1999yt,Dine:2003ax,Cline:2006ts,Canetti:2012zc}). Two of the most studied are electroweak baryogenesis \cite{Kuzmin:1985mm} and leptogenesis \cite{Fukugita:1986hr}. In the absence of CPT violation, such scenarios must satisfy Sakharov’s conditions \cite{Sakharov:1967dj}: baryon-number violation, C and CP violation and departure from thermal equilibrium. In electroweak baryogenesis, anomalous sphaleron processes provide baryon-number violation, while CP-violating sources, typically arising in extensions of the Standard Model, operate near the walls of bubbles during a first-order electroweak phase transition.  The bubble expansion supplies the required non-equilibrium environment. Leptogenesis creates a lepton asymmetry through the out-of-equilibrium, CP-violating decays of heavy Majorana neutrinos  that violate lepton number. Standard Model sphalerons then partially convert this lepton asymmetry into a baryon asymmetry. In contrast, in baryogenesis from primordial black holes (PBHs), the out-of-equilibrium condition is inherent in the Hawking evaporation and we consider the possibility that a higher-dimension operator coupling the Kretschmann scalar to a baryon-number–violating current can generate an effective chemical potential at the horizon, biasing the emitted Hawking flux. 

Alternative baryogenesis scenarios involving black holes were also explored by Hawking~\cite{Hawking:1974rv} and Carr~\cite{Carr:1976zz} in the 70s. In particular, they studied the evaporation of light PBHs into nucleons that subsequently decay asymmetrically producing more baryons than antibaryons. Baryogenesis via PBH evaporation has since attracted a lot of attention~\cite{Fujita:2014hha,Perez-Gonzalez:2020vnz,Bernal:2022pue,Hooper:2020otu,ShamsEsHaghi:2022azq,Gehrman:2022imk,Datta:2020bht}.
Recently, several authors~\cite{Hook:2014mla,Hamada:2016jnq,Boudon:2020qpo,Smyth:2021lkn} revisited the role of PBHs in the context of baryon asymmetry. Specifically, they studied a scenario where a chemical potential around the PBH is dynamically generated due to the presence of a higher order ($d>4$) operator, biasing Hawking evaporation towards baryon over antibaryon production. Several mechanisms can induce such an effect, including spontaneous baryogenesis~\cite{COHEN1987251,Cohen:1988kt},  gravitational
fluctuations produced during inflation \cite{Alexander:2004us} or the mass loss of an evaporating PBH \cite{Hamada:2016jnq}. In this work, we consider the final scenario, similar to Ref.~\cite{Hamada:2016jnq},
where it was  pointed out that as Hawking radiation steadily reduces the black hole mass,
\(M=M(t),\) the curvature invariants that are a function of \(M\), become time-dependent. In particular, let us consider the following dimension-eight, CP-violating operator  
\begin{equation}\label{eq:CPVop}
\frac{1}{M_{\star}^{4}}\,
\partial_\alpha\!\bigl(\mathcal R_{\mu\nu\rho\sigma}\mathcal R^{\mu\nu\rho\sigma}\bigr)\,
J^{\alpha}\,,
\end{equation}
where $M_{\star}$ is the cut-off scale of the new physics that induces the CP-violating coupling, $\mathcal R_{\mu\nu\rho\sigma}$ is the Riemann tensor and $J^{\alpha}$ is the baryon (or lepton) number–violating current.  
This operator is non–vanishing whenever the time derivative of the Kretschmann scalar,  
$\mathcal K\equiv\mathcal R_{\mu\nu\rho\sigma}\mathcal R^{\mu\nu\rho\sigma}$,  is non-zero. The Kretschmann scalar outside a PBH is given by
\begin{equation}
\mathcal K\equiv\mathcal R_{\mu\nu\rho\sigma}\mathcal R^{\mu\nu\rho\sigma}=\frac{3 M(t)^2}{4 \pi^2 M^4_{\mathrm{pl}}r^6}\,,
\end{equation}
where $r$ is the Schwarzschild radius,  \(M_{\mathrm{pl}}\equiv(8\pi G)^{-1/2}\simeq2.43\times10^{18}\,\mathrm{GeV}\)
is the reduced Planck mass with \(G\) being Newton's constant, and one finds that
\(
\dot{\mathcal K}\neq 0
\) whenever \(\dot M\neq 0\). This non-vanishing time derivative induces the operator $(\partial_{\alpha}\mathcal K)J^{\alpha}/M_{\star}^{4}$ to be non-zero and creates an effective CP-violating chemical potential at the black hole event horizon \cite{Hamada:2016jnq},
\be\label{eq:chempot}
\mu  = \frac{\partial_0\mathcal{K}}{M^4_*} \bigg\rvert_{r=r_{\rm S}} =-\frac{3}{2}(8 \pi)^6 \varepsilon(M) M_{\mathrm{pl}}\left(\frac{M_{\mathrm{pl}}}{M}\right)^7\left(\frac{M_{\mathrm{pl}}}{M_{\star}}\right)^4\,,
\ee
where \( \varepsilon(M_{\rm BH}) \) is the mass evaporation function of the black hole and $r_{\rm S}$ is the event horizon radius. The evaporation function encapsulates the total degrees of freedom available for emission at a given black hole mass, as described in Ref.~\cite{Cheek:2021odj}. The induced chemical potential grows as the black hole evaporates and produces a baryon (or lepton) asymmetry.
In Ref.~\cite{Hamada:2016jnq}, the authors find a viable region in parameter space where the evaporation of light PBHs can account for the observed baryon asymmetry. However, the results are obtained under a set of simplifying assumptions. First, the authors only consider the evaporation of PBHs into spin $0$ and spin $1/2$ particles, and second, they neglect the entropy dilution coming from the evaporation into photons. Third, the dependence of the greybody factors on the effective chemical potential is also neglected, and finally, the baryonic yield is obtained from an approximate analytic treatment rather than a full solution of the coupled evolution equations for an evaporating PBH population in an expanding universe. 

In this work, we build on the results of Ref.~\cite{Hamada:2016jnq}, by going beyond the approximated analytical analysis and investigating the interplay between this CP-violating operator and PBHs via numerical simulations. Our paper is laid out as follow: we review the physics of PBHs in Sec.~\ref{sec:PBHs} and provide some analytical understanding of the lepton asymmetry generated from a PBH in Sec.~\ref{sec:LAPBH}. In Sec.~\ref{sec:BEs}, we compare the analytical approach with our numerical setup and present our results and conclusions in Secs.~\ref{sec:results} and \ref{sec:conclusions}, respectively. In Appendix~\ref{sec:AppGBfactors}, we provide the details of the computation of the greybody factors including the chemical potential, which we use in our numerical simulations.

\section{Primordial Black Holes}\label{sec:PBHs}
We first review four representative BH mass spectra: monochromatic, log-normal, power-law, and (generalized) critical-collapse distributions in Sec.~\ref{sec:fms} and follow in Sec.~\ref{sec:evaporation} by detailing the evaporation of microscopic Schwarzschild BHs.

\subsection{Formation and Mass Spectrum}\label{sec:fms}

Primordial black holes can form in the early Universe through the collapse of overdensities generated by mechanisms such as inflation~\cite{Ivanov:1994pa,Randall:1995dj,Garcia-Bellido:1996mdl}, first-order phase transitions~\cite{Baker:2021sno,Gross:2021qgx,Kawana:2021tde,Liu:2021svg,Baldes:2023rqv,Flores:2024lng,Gouttenoire:2023naa,Lewicki:2023ioy}, or topological defects~\cite{Hawking:1987bn,Polnarev:1988dh,Brandenberger:2021zvn,Ipser:1983db,Ferrer:2018uiu,Liu:2019lul,Gouttenoire:2023gbn,Ge:2023rrq,Lu:2024ngi}. 
We consider a population of Schwarzschild PBHs formed during the radiation-dominated era, potentially characterized by an extended initial mass distribution, \( dn_{\rm PBH}/dM \). We define the normalized initial number distribution as
\begin{align*}
    \fBH(M) = \frac{1}{n_{\rm PBH}} \frac{dn_{\rm PBH}}{dM}\,,
\end{align*}
such that it satisfies the normalization condition
\begin{align}
    \int_{0}^{\infty} \fBH(M)\, dM = 1\,.
\end{align}
From this distribution, the initial comoving PBH energy density is given by
\begin{align}
    \rho_{\rm BH}^{\rm in} = n_{\rm PBH}^{\rm in} \int_0^{\infty} M\, \fBH(M)\, dM\,,
    \label{eq:def_rhoBH}
\end{align}
where \( n_{\rm PBH}^{\rm in} \) denotes the PBH number density at formation. This number density can be related to the properties of the primordial plasma.
We assume that the distribution \( \fBH(M) \) is centered around a characteristic mass scale \( M_{\rm scl} \), which we associate with the mass contained within the particle horizon at the time of formation, corresponding to a radiation temperature \( T_0 \)
\begin{equation}
    M_{\rm scl} \equiv M_{\rm scl}(T_0) = \frac{4\pi}{3} \kappa \frac{\rho_R(T_0)}{H^3(T_0)}\,,
    \label{eq:horizon-mass}
\end{equation}
where \( \kappa \simeq 0.2 \) parametrizes the efficiency of collapse, \( \rho_R(T_0) \) is the radiation energy density, and \( H(T_0) \) is the Hubble parameter at temperature \( T_0 \).
Using Eq.~\eqref{eq:horizon-mass}, the formation temperature for a PBH of mass \( M_{\rm scl} \) is approximately
\begin{equation}
    T_0 \simeq 4.3 \times 10^{15}\, \mathrm{GeV}
    \left( \frac{106.75}{g_*} \right)^{1/4}
    \left( \frac{1~\mathrm{g}}{M_{\rm scl}} \right)^{1/2}
    \left( \frac{\kappa}{0.2} \right)^{1/2},
    \label{eq:T0}
\end{equation}
where \( g_* \) is the effective number of relativistic degrees of freedom at the time of formation. This implies that gram-scale PBHs produced from inflationary fluctuations must have formed at extremely high temperatures, \( T_0 \gtrsim 10^{15}\, \mathrm{GeV} \).

The fractional contribution of PBHs to the total energy density at formation is defined as
\begin{equation}
    \beta \equiv \frac{\rho_{\rm BH}^{\rm in}(T_0)}{\rho_R(T_0)}.
\end{equation}
The exact form of the PBH mass distribution \( \fBH(M) \) depends on the specific production mechanism.
We will consider the following four well-known cases for the normalized mass distribution.
\begin{itemize}

    \item {\bf Monochromatic}.
        A simple scenario corresponds to consider that all PBH were generated with the same mass $M_{\rm scl}$ such that
        \begin{align}
            \fBH(M) = \delta(M-M_{\rm scl})\,.
        \end{align}
        Although unrealistic, this monochromatic scenario is helpful to understand the main features of PBH evaporation and its generation of baryon asymmetry.

    \item {\bf Log-Normal (LN)}.
        The formation of PBHs during inflation typically requires a brief phase of \emph{ultra-slow-roll} dynamics, which amplifies the primordial power spectrum of scalar curvature perturbations and generates a localized peak~\cite{Ballesteros:2017fsr, Karam:2022nym, Dalianis:2018frf, Heurtier:2022rhf}. This enhancement generally results in a log-normal PBH mass function~\cite{Dolgov:1992pu}, a feature that has been confirmed both numerically and analytically in the context of slow-roll inflation~\cite{Dolgov:2008wu, Green:2016xgy}. The resulting mass distribution takes the form
        \begin{align}
            \fBH(M) = \frac{1}{\sqrt{2\pi}\sigma M}\exp\left[-\frac{\log^2(M/M_{\rm scl})}{2\sigma^2}\right],
            \label{eq:distri_lognormal}
        \end{align}
        where $M_{\rm scl}$ denotes the central mass of the distribution and $\sigma$ characterizes its width. 

    \item{\bf Power-Law (PL)}.
        An alternative mechanism for PBH formation involves the collapse of a nearly scale-invariant spectrum of primordial curvature perturbations in a Universe dominated by a perfect fluid with a constant equation-of-state parameter $w$~\cite{Carr:1975qj}. In this scenario, the resulting PBH mass distribution follows a power-law form,
        \begin{equation}
            \fBH(M)\propto 
            \begin{cases}
                M^{-\alpha}, & \text{for } M_{\rm scl} \leq M \leq M_{\rm scl} \times 10^{\sigma}\,, \\
                0,           & \text{otherwise}\,.
            \end{cases}
            \label{eq:distri_powerlaw}
        \end{equation}
        where the exponent $\alpha$ is determined by
        \begin{equation}
            \alpha \equiv \frac{4w + 2}{w + 1}\,.
        \end{equation}
        The mass range $[M_{\rm scl},\, M_{\rm scl} \times 10^\sigma]$ is set by the range of scales over which the scale-invariant spectrum extends. In physically relevant scenarios where the Universe is no longer inflating, the equation-of-state parameter typically lies in the range $-1/3 < w \leq 1$, which corresponds to
        \begin{equation}
            1 < \alpha \leq 3\,.
            \label{eq:range}
        \end{equation}

    \item{\bf Critical Collapse (CC)}.
        The application of critical scaling to gravitational collapse provides a more refined description of PBH formation from primordial density fluctuations~\cite{Choptuik:1992jv,Yokoyama:1998qw,Yokoyama:1998xd,Kuhnel:2015vtw}. While early models assumed that PBHs form with masses equal to the horizon mass at the time of collapse, critical phenomena reveal a more complex picture: there exists an upper cutoff near the horizon mass, but the mass distribution extends to smaller values, forming a characteristic low-mass tail. This behavior has been shown to arise generically across a wide range of inflationary models~\cite{Kuhnel:2015vtw}. The resulting PBH mass function takes the form
        \begin{equation}
            \fBH(M) \propto M^{1.25}\exp\left[-\left(\frac{M}{M_{\rm scl}}\right)^{2.85}\right]\,,
        \end{equation}
        where $M_{\rm scl}$ denotes the characteristic mass scale of the distribution.
        However, one can generalize the CC mass function to better reproduce the PBH mass distribution generated by several representative primordial curvature spectr, see Ref~\cite{Gow:2020cou}.
        Such generalized critical collapse (GCC) mass function is parametrized as
        \begin{align}
            \fBH(M) = \frac{\eta}{\mu\, \Gamma(\kappa/\eta)} \left(\frac{M}{\mu}\right)^{\kappa - 1}\exp\left[-\left(\frac{M}{\mu}\right)^\eta\right],
        \end{align}
        with $\eta >0, \kappa>1$~\cite{Gow:2020cou, Klipfel:2025jql} are the parameters of the GCC distribution, such that it peaks at $M_{\rm scl} = \mu [(\kappa-1)/\eta]^{1/\eta}$.
\end{itemize}
Having considered the initial distribution of our PBH population, we can now focus on its possible evaporation in the Early Universe.

\subsection{Evaporation and Hawking Radiation}\label{sec:evaporation}
The Schwarzschild radius, \( r_{\mathrm{S}} \), of a light black hole can be microscopic,
\begin{equation}
    r_{\mathrm{S}} \sim 2.7 \times 10^{-5}\left(\frac{M}{10^{10}\,\mathrm{g}}\right)\mathrm{fm}\,,
\end{equation}
and when this radius becomes comparable to the Compton wavelength of a particle species, the PBH can emit particles of that species as Hawking radiation. The temperature associated with this Hawking radiation is inversely proportional to the black hole mass:
\begin{equation}
    T_{\mathrm{BH}} =\frac{M_{\mathrm{pl}}^{2}}{M}\simeq 1~\mathrm{TeV}\left(\frac{10^{10}\,\mathrm{g}}{M}\right)\,.
    \label{eq:T_BH}
\end{equation}
As a black hole evaporates it converts its mass into a spectrum of Hawking quanta.  
The differential emission rate, per unit time and per unit momentum, for a particle species \(i\) of mass \(m_i\), spin \(s_i\), internal degrees of freedom \(g_i\) and chemical potential \(\mu_i\) is
\begin{equation}\label{eq:rateinsta}
\frac{d^2 n_i}{dp \, dt} =\frac{g_i}{2 \pi^2} \frac{\sigma_{s_i}\left(M,p, \mu_i\right)}{\exp \left[\left(E_i(p)+\mu_i\right) / T_{\mathrm{BH}}\right]-(-1)^{2 s_i}} \frac{p^3}{E_i(p)}\,.
\end{equation}
The cross-section \(\sigma_{s_i}(M,p,\mu_i)\) is often recast as the greybody factor:
\begin{equation}
\Gamma_{s_i}\equiv \frac{\sigma_{s_i} p^{2}}{\pi},
\end{equation}
which quantifies how the gravitational and centrifugal barriers surrounding the black hole partially reflect outgoing quanta, thereby modifying the Hawking spectrum~\cite{Hawking:1974rv,Hawking:1974sw,Page:1976df,Page:1977um}.
\begin{figure}[t!]
  \centering
  \includegraphics[width=0.65\linewidth]{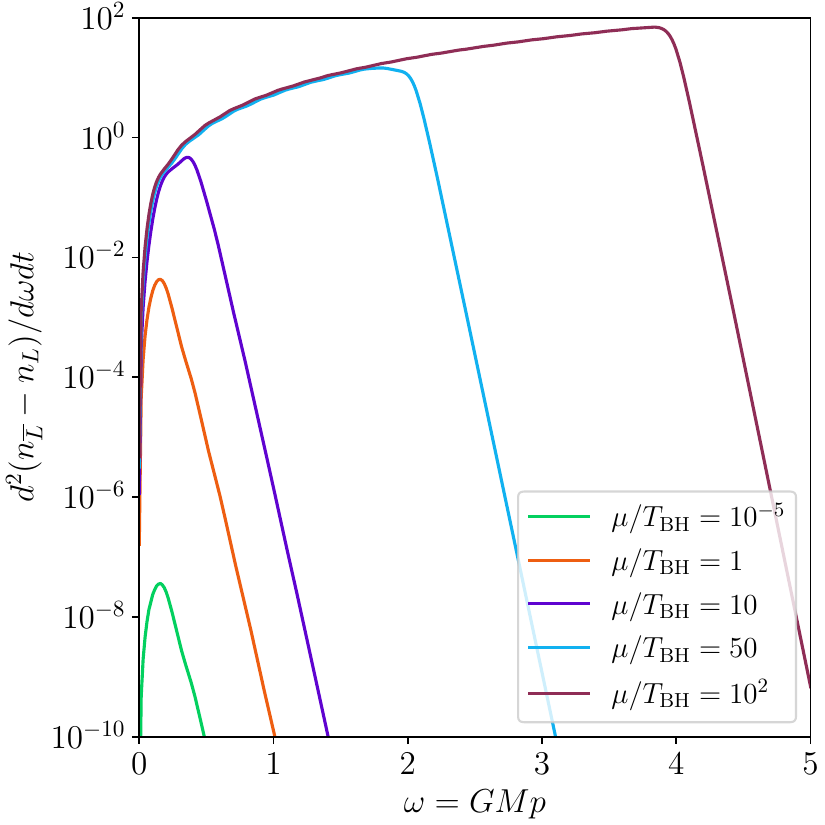}
  \caption{Differential emission rate asymmetry for leptons for several values of the ratio $\mu/T_{\rm BH}$.}
\label{fig:dndEdt}
\end{figure}
In \figref{fig:dndEdt} we show the difference of the differential emission rate between leptons and antileptons. As expected, in the limit where the chemical potential is negligible ($\mu/T_{\rm BH} \ll 1$), the differential rate for both leptons and antileptons is roughly the same, or, in other words, Hawking evaporation is symmetric. Therefore, the differential emission rate asymmetry is negligible. On the other hand, as the value of $\mu/T_{\rm BH}$ becomes significant, one can see that Hawking evaporation yields more antileptons than leptons. The large suppression at large values of $\omega$ comes from the exponential function in \equaref{eq:rateinsta}.

Integrating the differential rate of \equaref{eq:rateinsta} over momentum and summing over all particle species \(i\) (for simplicity we also neglect the chemical potential), one finds the net mass loss is~\cite{Hawking:1975vcx,MacGibbon:1990zk,MacGibbon:1991tj}
\begin{equation}\label{eq:ml}
\frac{dM}{dt} = -\sum_i \int_{m_i}^{\infty} \frac{d^{2} n_i}{dt \, dE}\, E\, dE = -\frac{(8\pi)^2 \varepsilon(M) M_{\mathrm{pl}}^{4}}{M^{2}}\,,
\end{equation}
where \(t\) denotes time and \(M\) the instantaneous black hole mass.  
Let us note that in order to obtain the mass lose rate of Eq.~\eqref{eq:ml}, we have assumed the validity of the semi-classical approximation, in which quantum fields propagate on a fixed classical spacetime, that is, neglecting backreaction effects\footnote{The ``memory burden'' hypothesis~\cite{Dvali:2024hsb} proposes late-time deviations from the standard mass loss rate, motivated by analogies with solitonic configurations in non-Abelian gauge theories and assuming the N-graviton portrait of a BH. As it remains speculative, we do not include these effects in our analysis.}.
This treatment is expected to fail near the Planck scale, where the adiabaticity condition $\dot{T}_{\rm BH}/T_{\rm BH}^2 \ll 1$ breaks down~\cite{Barcelo:2010pj}.
To remain conservative, we will consider the effect of stopping the evolution at $M =M_{\rm pl}$ and $M \approx 100\,M_{\rm pl}$, showing the effect on the generated baryon asymmetry.
Further complications arise from the information loss problem. After the \emph{Page time}~\cite{Page:1993wv,Page:2013dx,Perez-Gonzalez:2025try}, the von Neumann entropy of Hawking radiation exceeds the Bekenstein-Hawking entropy, suggesting potential non-thermal corrections to the emission spectrum. As the PBHs of interest have long surpassed this point, such effects could in principle modify the spectrum~\cite{Perez-Gonzalez:2025try}. However, in the absence of a well-established framework for these corrections~\cite{Buoninfante:2021ijy,Buoninfante:2025gqk}, we assume Eq.~\eqref{eq:ml} as valid throughout the PBH lifetime until near the Planck scale.

Integrating the mass loss rate of Eq.~\eqref{eq:ml} from the initial mass \(M_{\mathrm{in}}\) down to \(100 \cdot M_{\rm pl}\) gives the total lifetime of a PBH,
\begin{equation}\label{eq:lifetime}
  \tau \;\simeq\; \frac{160}{\pi g_*}\,
  \frac{M_{\mathrm{in}}^{3}}{M_{\mathrm{pl}}^{4}} \, ,
\end{equation}
where we assumed that $M_{\mathrm{in}}\gg100 \cdot M_{\rm pl}$ and $\varepsilon(M) \simeq g_*/30720\pi$. 
Using this lifetime and assuming the Universe continues to be radiation dominated, we can compute the radiation temperature when the PBH population completely evaporates~\cite{Bernal:2022pue}:
\begin{equation}\label{eq:tev_rad}
  T_{\mathrm{ev}}^{(\mathrm{rad})} 
  \;\simeq\; 1.2\times10^{10}\,\mathrm{GeV}\,
  \left(\frac{g_*}{106.75}\right)^{1/4}
  \left(\frac{1\,\mathrm{g}}{M_{\mathrm{in}}}\right)^{3/2}\,.
\end{equation}
If, on the other hand, PBH evaporation occurs when the Universe’s energy density is PBH-dominated, 
the radiation temperature at evaporation now depends on the PBH’s abundance:
\begin{equation}\label{eq:tev_pbh}
  T_{\mathrm{ev}}^{(\mathrm{PBH})} 
  \;\simeq\; 4.1\times10^{10}\,\mathrm{GeV}\,
  \left(\frac{g_*}{106.75}\right)^{5/12}
  \left(\frac{10^{-3}}{\beta}\right)^{1/3}
  \left(\frac{0.2}{\kappa}\right)^{1/6}
  \left(\frac{1\,\mathrm{g}}{M_{\mathrm{in}}}\right)^{11/6}.
\end{equation}
Both \equasref{eq:tev_rad}{eq:tev_pbh} are approximations that describe the Standard Model plasma just before the PBHs evaporate suddenly.

\section{Lepton asymmetry generated from a PBH}\label{sec:LAPBH}
An \emph{unbiased} black hole emits leptons via Hawking radiation at a rate determined by the lepton’s internal degrees of freedom and the energy‐dependent greybody factor (see \emph{e.g.} Eq.~(\ref{eq:rateinsta})).  Consequently, antileptons are produced at the same rate and the net lepton number vanishes. 
However, the presence of the curvature–induced operator of Eq.~\eqref{eq:CPVop} creates a
local chemical potential \(\mu(M)\) that can favor the production of leptons over anti-leptons, or vice versa. To quantify the bias, we define the following lepton asymmetry rate: 
\be\Gamma\equiv
    \int dp\!\left[
  \frac{d^{2} n_{\overline{L}}}{dp\,dt}
  -\frac{d^{2} n_{L}}{dp\,dt}
\right]
= \frac{8\pi\,M_{\mathrm{pl}}^{2}}{M}
  \int d\omega\!\left[
    \frac{d^{2} n_{\overline{L}}}{dp\,dt}
    -\frac{d^{2} n_{L}}{dp\,dt}
  \right]
= \frac{8\pi\,M_{\mathrm{pl}}^{2}}{M}\,
  \widetilde{\Gamma}_{\rm BH\to\Delta L}\,,
\label{gammadef}
\ee
where we have rescaled the momentum to the dimensionless variable  
\(\omega \equiv Mp/(8\pi M_{\mathrm{pl}}^{2})\).  In our numerical analysis, we work with the dimensionless rate \(\widetilde{\Gamma}_{\rm BH\to\Delta L}\).
We will now consider the lepton asymmetry rate in several limits. First, we will assume  that 
 \(\mu\ll T_{\mathrm{BH}}\) and that the absorption cross-section for particle and antiparticle are identical and momentum independent such that $\sigma_{s_i}\left(M,p, \mu\right) = \sigma_0$.  To first order in $\mu/T_{\rm BH}$, we find that the lepton asymmetry rate is
\be
\Gamma \approx-\frac{g_L \sigma_0 \mu T_{\mathrm{BH}}^2}{6}\,,
\ee
and hence the lepton asymmetry, increases linearly with $\mu$.
The second analytic limit we consider is that simply the absorption cross-section is momentum and chemical potential independent and we find that the lepton asymmetry rate can be approximated as 
\be\label{eq:lepass}
\Gamma\approx 
 -\sigma_0\frac{g_L T^3_{\rm BH}}{\pi^2}\left( \operatorname{Li}_3(-e^{-\mu/T_{\rm BH}}) - \operatorname{Li}_3(-e^{\mu/T_{\rm BH}})
\right)\,.
\ee
For large $\mu/T_{\rm BH}$, the polylogarithms grow
approximately as \((\mu/T_{\mathrm{BH}})^{3}/6\) and therefore the asymmetry 
rises faster than linearly, $\Gamma \propto \mu^3$, at large chemical potential. 
Nonetheless, for a more accurate estimation of the lepton asymmetry from evaporation, 
we  compute the greybody factors, including the dependence on the chemical potential $\sigma_{1 / 2}(M, p,\mu)$ and without relying on any assumption for a wide range of values for $\mu/T_{\rm BH}$ (see discussion in Appendix~\ref{sec:AppGBfactors}). 

\begin{figure}
    \centering
    \includegraphics[width=0.65\textwidth]{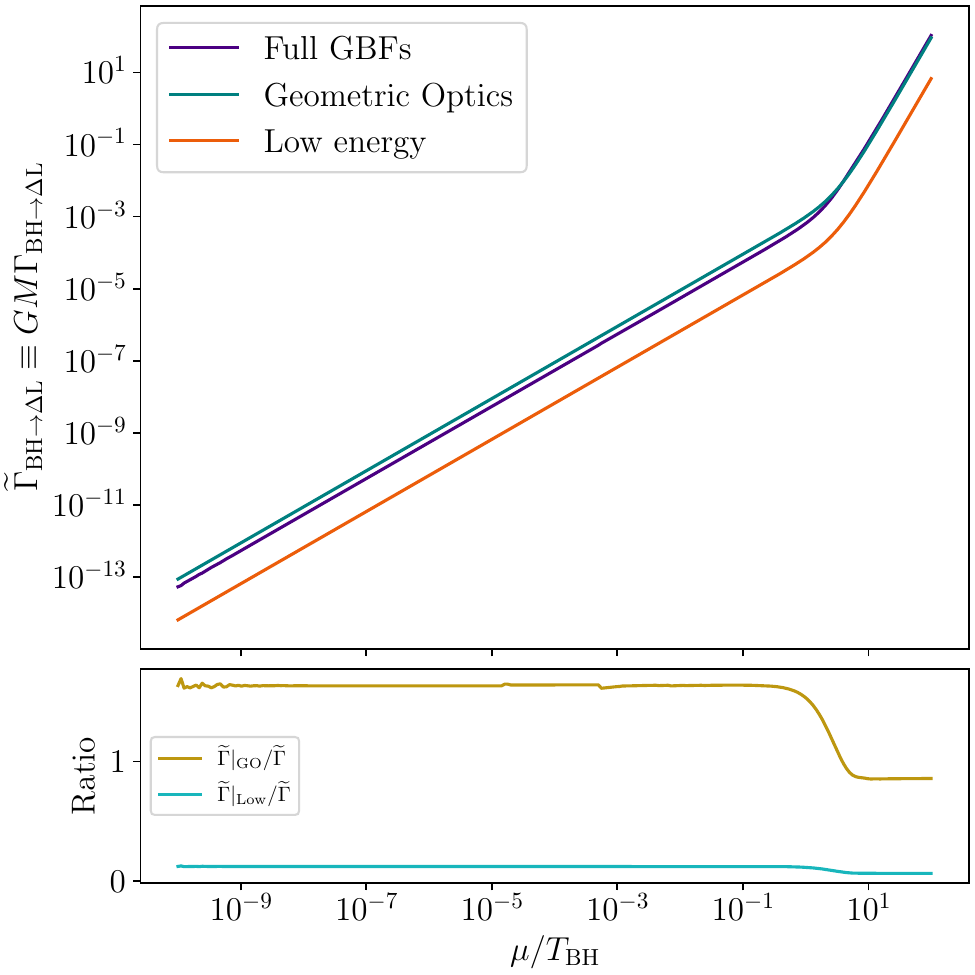}
    \caption{
    Top: Dependence of the dimensionless emission rate $\widetilde{\Gamma}_{\rm BH\to\Delta L}$ on $\mu/T_{\rm BH}$ obtained by considering the geometric optics (green), low energy (orange) limits for the greybody together with greybody factors computed including the dependence on the chemical potential.
    Bottom: Ratio of the two limiting cases, geometric optics (olive) and low energy (light blue), to the exact result.}
    \label{FigGammaBH}
\end{figure}

We compare the lepton asymmetry production rate for different greybody factor prescriptions in \figref{FigGammaBH}. Specifically, we show the dependence of the dimensionless rate $\widetilde{\Gamma}_{\rm BH\to\Delta L}$, assuming for $\sigma_0$ the geometric optics limit $EM/M_\mathrm{pl}^2\gg1$ (green), the low-energy approximation $EM/M_\mathrm{pl}^2\ll1$ (orange), and the full numerical greybody factors (purple). The bottom panel displays the ratio of the two limiting cases, geometric optics (olive) and low energy (light blue), to the exact result.
All three prescriptions exhibit the same qualitative behavior: for $\mu/T \gg 1$, the rate scales as $(\mu/T_{\rm BH})^3$, while for $\mu/T_{\rm BH} \ll 1$, it scales linearly with $\mu/T_{\rm BH}$. As expected, the two limiting cases, which assume $\sigma_0$ to be independent of $\mu/T_{\rm BH}$, differ from each other by an overall normalization. 
However, the full rate shows a slightly different dependence on $\mu/T_{\rm BH}$. 
In the $\mu/T_{\rm BH} \ll 1$ regime, the geometric optics approximation overestimates the correct rate due to its significant overestimation of the greybody factors.
This discrepancy diminishes near $\mu/T_{\rm BH} \sim 1$, where contributions from higher angular momentum modes become relevant. As seen in Fig.~\ref{fig:dndEdt}, for some range of momenta $p$, the condition $\mu/T_{\rm BH} > E_i(p)/T_{\rm BH}$ is satisfied, reducing the exponential suppression typically affecting higher-$\ell$ modes in the absence of a chemical potential. 
As a result, the exact rate exceeds the geometric optics and low-energy approximations by a factor of approximately 1.15 and 15, respectively.
This highlights the critical importance of using accurate greybody factors when evaluating lepton asymmetry production in this context. 
In the following sections, we use the exact greybody factors for our study of lepton asymmetry production.

\section{Friedmann and Boltzmann Equations for PBH Baryogenesis}\label{sec:BEs}
In this section, we begin in Sec.~\ref{sec:antreatment} by reviewing the analytic treatment on how a baryon asymmetry can be generated from a population of biased PBHs and follow in Sec.~\ref{sec:numtreatment} by presenting our numerical treatment of the problem.

\subsection{Analytic Treatment}\label{sec:antreatment}
To compute the lepton asymmetry produced by a population of PBHs in the early Universe, given the CP- and lepton number $L$-violating operator of \equaref{eq:CPVop}, one must take into account the complete set of coupled equations describing both PBH and radiation energy density evolution in an expanding Universe. This level of detail was not fully addressed in Ref.~\cite{Hamada:2016jnq}, which instead approximated the lepton asymmetry as:
\begin{equation}
\frac{n_L}{s} = \frac{n_{\rm PBH}}{s}\,\delta N_L = \Omega_{\rm PBH}\,\frac{\rho_{\rm tot}}{s}\,\frac{\delta N_L}{M_{\rm ev}} 
= \frac{3\,\Omega_{\rm PBH}T_{\rm ev}\,\delta N_L}{4\,M_{\rm ev}}\,,
\label{leptasym0}
\end{equation}
where $\Omega_{\rm PBH}$ is the ratio of the PBH energy density to the total energy density at the time of evaporation (when the Universe’s temperature is $T_{\rm ev}$), $M_{\rm ev}$ is the PBH mass at evaporation, and $\rho_{\rm tot}/s = 3\, T_{\rm ev}/4$. The quantity \(\delta N_L\) estimates the lepton asymmetry generated by a single black hole and is obtained by integrating the lepton asymmetry rate relative to the radiated energy carried per each type of emitted particle over the lifetime of the black hole:
\begin{equation}
\delta N_L = \int_{M_{\rm min}}^{M_{\rm PBH}^0} dM \frac{\dot{n}_L - \dot{n}_{\bar{L}}}{\int dp\, p \frac{d^2 n_L}{dpdt} + \int dp\, p \frac{d^2 n_{\bar{L}}}{dpdt}+ \int dp\, p \frac{d^2 n_{\rm other}}{dpdt}}\,.
\label{deltaNLestimate1}
\end{equation}
Changing the integration variable from $M$ to $X=(M/M_{\rm cr})^2$ (see \equaref{Mcriticaldef}), we can rewrite \equaref{deltaNLestimate1} as
\begin{equation}
\delta N_L = \frac{1}{2} \left( \frac{M_{\rm cr}}{M_{\mathrm{pl}}} \right)^2 \int^{X_0}_{X_{\rm min}} dX g(X)\,.
\label{deltaNLestimate2}
\end{equation}
Note that the value of $X_{\rm min}$ will be in general different from zero because as the PBH evaporates both the black hole temperature and chemical potential increase and will eventually reach the cutoff scale $M_{\star}$, beyond which our effective field theory breaks down. Therefore, $X_{\rm min}$ will be given by the following condition
\begin{equation}\label{eq:breakdown}
M_{\min }= \max \!\Biggl(
\frac{M_{\mathrm{pl}}^2}{M_{\star}}\,,\,
\biggl(\frac{M_{\mathrm{cr}}^6\,M_{\mathrm{pl}}^2}{M_{\star}}\biggr)^{\!1/7}
\Biggr)\,,
\end{equation}
where the critical mass $M_{\mathrm{cr}}$ is set by the condition $\mu \sim T_{\rm BH}$, which corresponds to the point in the black hole’s evolution at which the lepton asymmetry reaches its maximum. This quantity is defined to be
\begin{equation}\label{eq:mcritical}
M_{\mathrm{cr}}\sim 10 \,\biggl(\frac{M_{\mathrm{pl}}}{M_{\star}}\biggr)^{\!\!2/3} M_{\mathrm{pl}}\,.
\end{equation}
One can check that $X_{\rm min}=1$ for $M_{\star}=10^{-3} M_{\rm pl}$, for which both the temperature and chemical potential reach $M_{\star}$ at the same time. For $M_{\star}<10^{-3} M_{\rm pl}$ ($X_{\rm min}>1$), it is the PBH temperature that sets the limit, while for $M_{\star}>10^{-3} M_{\rm pl}$ ($X_{\rm min}<1$), it is the chemical potential that will reach the cutoff scale first. As discussed in Ref.~\cite{Hamada:2016jnq}, this provides a conservative estimate of the lepton asymmetry, since, in principle, the PBH would still evaporate asymmetrically beyond this point.

\begin{figure}[t]
    \centering
    \includegraphics[width=1\textwidth]{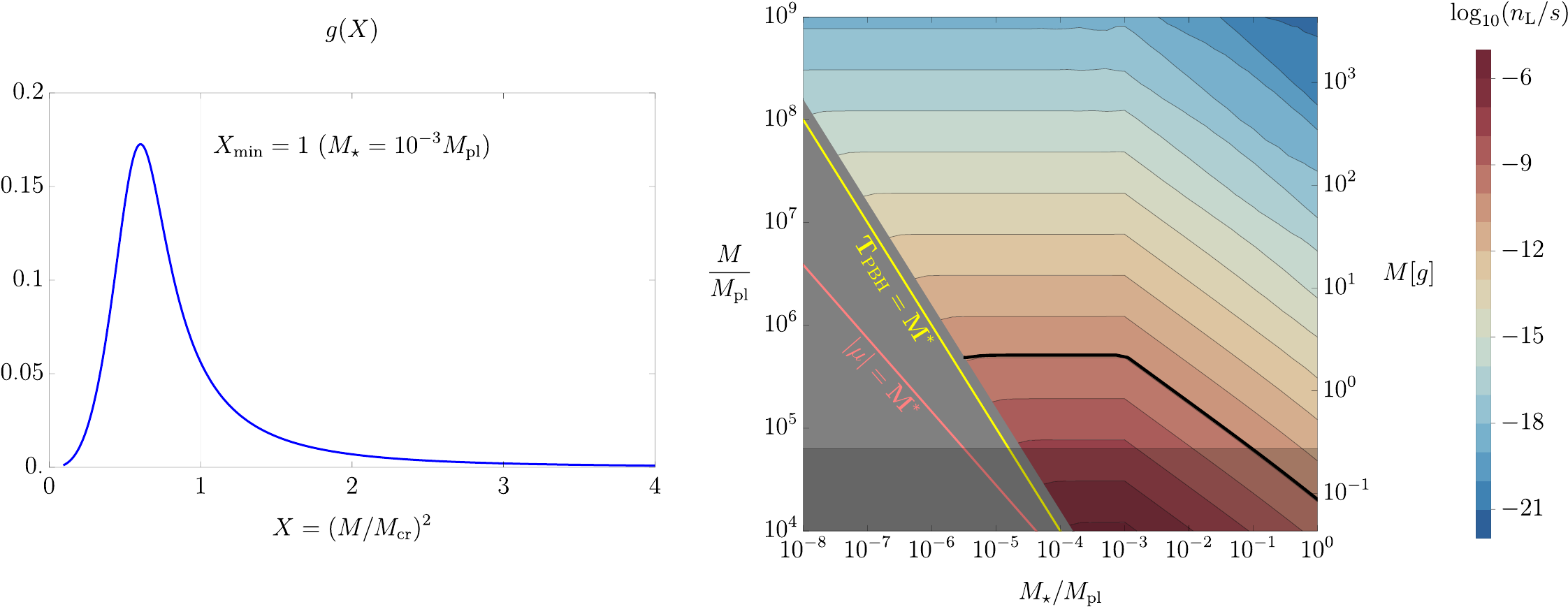}
    \caption{The function $g(X)$ defined in \equaref{deltaNLestimate2} (left) and the resulting lepton yield $n_L/s$ for different values of the cutoff scale and PBH initial mass in units of the reduced Planck mass for $\Omega_\mathrm{PBH}=1$ (right). The grey shaded bottom left corner indicates the region where the PBHs are initially formed with a temperature (below the yellow line) and/or a chemical potential (below the pink line) above $M_{\star}$. The horizontal shaded band is ruled out by the tensor-to-scale ratio as discussed in the text. The solid thick line indicates the observed lepton yield.}
    \label{fig1}
\end{figure}

The behavior of \(g(X)\) is shown in the left panel of \figref{fig1}. We see that $g(X)$ peaks at around $X \sim 0.6$, or, equivalently, $M \sim 0.8 M_{\rm cr}$. This is the point when the PBH most efficiently produces a lepton asymmetry. As explained in Ref.~\cite{Hamada:2016jnq}, for larger values of $X$ the chemical potential becomes small, and Hawking evaporation is essentially symmetric, whereas for smaller values, although the bias in Hawking evaporation is maximal, the emitted particles (in this case, leptons or antileptons) carry a lot of energy. 

In deriving \equaref{leptasym0} and \equaref{deltaNLestimate1}, it is assumed that PBHs only evaporate into massless scalars and fermions, using the low-energy limit for the greybody factors. Moreover, entropy dilution, which is important for estimating both dark matter production and the baryon asymmetry in PBH scenarios~\cite{Perez-Gonzalez:2020vnz,Cheek:2021cfe,Calabrese:2023key,Bernal:2022pue}, is neglected. Following the procedure in Ref.~\cite{Hamada:2016jnq}, we reproduce the results shown in \figref{fig1} right, which shows the resulting lepton yield (see colour bar) as a function of the cutoff scale $M_{\star}$ and initial PBH mass $M$ for $\Omega_{\rm PBH}=1$. The grey triangle in the lower left corner indicates the region of parameter space where PBHs have either a temperature (below the yellow line) or a chemical potential (below the pink line) above $M_{\star}$ at formation, and hence the effective field theory is immediately invalid. The grey-shaded band highlights regions of parameter space excluded by inflation\footnote{This bound only applies to PBH formation mechanisms related to inflation, such as the collapse of large overdensities upon horizon reentry. However, we note that effects such as critical collapse can lead to arbitrarily light PBHs. In any case, we do not assume any particular formation mechanism and consider that PBHs above the Planck mass form with a mass distribution of the kind discussed in Section~\ref{sec:fms}.}~\cite{Masina:2020xhk, Baldes:2020nuv}.
Note that there is a clear change of behavior at $M_{\star}\simeq 10^{-3}M_{\mathrm{pl}}$. This is due to the integral cutoff $X_{\rm min}$ in \equaref{deltaNLestimate2}. For a fixed PBH mass, as we decrease the cutoff scale $M_{\star}$, the point of maximum lepton asymmetry production is reached earlier (see \equaref{eq:mcritical}). At the same time, from \equaref{eq:breakdown}, the time at which our effective theory breaks down is also reached earlier. Therefore, two opposing effects are at play as we decrease the value of $M_{\star}$. Due to the different dependence of the breakdown point on $M_{\star}$, we observe two different behaviors above and below $M_{\star}\simeq 10^{-3}M_{\mathrm{pl}}$.

\begin{figure}
    \centering
    \includegraphics[width=1\textwidth]{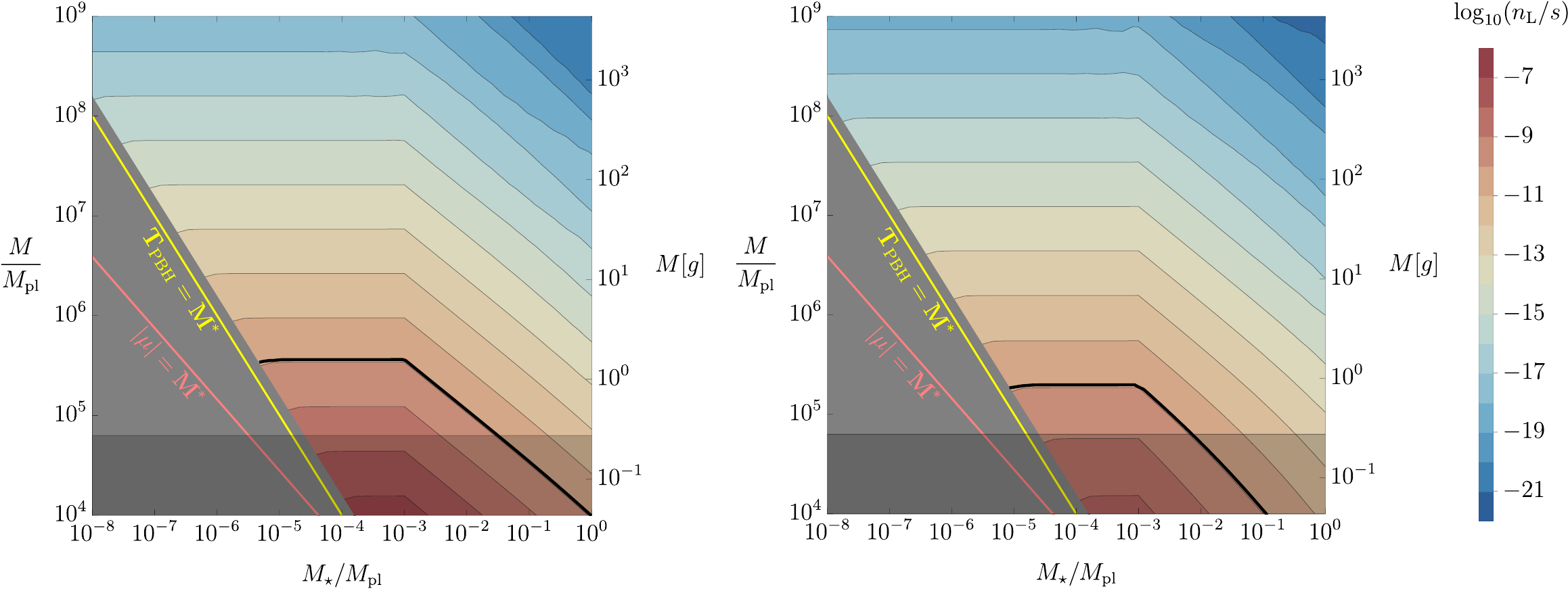}
    \caption{Lepton yield $n_L/s$ accounting for entropy dilution for $\beta=10^{-3}$ (left) and $\beta=10^{-5}$ (right). See \figref{fig1} for a description of the figure.}
    \label{fig2}
\end{figure}

We can now estimate the impact of entropy dilution by following Ref.~\cite{Bernal:2022pue}. Using energy conservation before and after PBH evaporation, we find that
\begin{equation}
    \frac{\pi^2}{30}g_\star T^4_{\rm ev}+M_{\mathrm{in}} n^0_{\rm PBH} \frac{s(T_{\rm ev})}{s(T_0)} = \frac{\pi^2}{30}g_\star \tilde{T}^4 \implies \frac{s(\tilde{T})}{s(T_{\rm ev})}=\left( \frac{\tilde{T}}{T_{\rm ev}} \right)^3 = \left( 1+\frac{\beta T_0}{T_{\rm ev}} \right)^{3/4},
    \label{eq:entropydil}
\end{equation}
where $n^0_{\rm PBH}=n_{\rm PBH}(T_0)$ is the initial PBH number density, $\tilde{T}$ is the temperature of the SM plasma after PBH evaporation and \(\beta = \rho_{\rm PBH}(T_0)\!/\!\rho_{\rm tot}(T_0)\), with \(\rho_{\rm PBH}(T_0)\) and \(\rho_{\rm tot}(T_0)\) denoting the energy densities of the PBHs and the total energy of the Universe at formation, respectively. We also define all relevant quantities at PBH formation time since the PBH mass is constant for most of its evolution and $n_{\rm PBH}(T_{\rm ev})/s(T_{\rm ev}) \simeq n_{\rm PBH}(T_0)/s(T_0)$. 
Applying the entropy dilution factor, the estimated lepton asymmetry is
\begin{equation}
\frac{n_L}{s} \;=\; \frac{n_{\rm PBH}}{s}\,\delta N_L \;=\; \frac{3\,\beta \,T_0\,\delta N_L}{4\,M_{\rm PBH}^0}
\left(1 + \beta\,\frac{T_0}{T_{\rm ev}}\right)^{-\frac{3}{4}},
\label{eq:lepasymdil}
\end{equation}
where we used \(\rho_{\rm tot}/s = 3\,T_0/4\). We present the resulting leptonic yield in \figref{fig2}. Although the inclusion of entropy dilution suppresses the lepton asymmetry relative to \equaref{leptasym0}, there is an additional factor of $T_0 / T_{\rm ev}$ that partially compensates for this suppression. We observe a similar behavior than in \figref{fig1}.

\subsection{Numerical Solution}\label{sec:numtreatment}

Having gained this insight, we now move on to a more detailed numerical treatment of the interplay between PBHs and the radiation component, closely following the approach in Refs.~\cite{Perez-Gonzalez:2020vnz,Cheek:2022mmy}. 
First, we write down the mass loss rate for the PBH population,
\begin{align}\label{eq:BoltzmannPBHandSM}
\frac{d \varrho_{\rm BH}}{d a} &= \frac{n_{\mathrm{BH}}^{\rm in}}{aH}\int\frac{dM }{d t}\, f_{\rm PBH} (M_{\rm in}^t)\,\frac{dM_{\rm in}^t}{dM} \,d M\,,
\end{align}
where $\varrho_{\mathrm{BH}} \equiv a^{3}\,\rho_{\mathrm{BH}}$ is the comoving energy density of the PBHs, $a$ is the scale factor (initialized to $a_0 = 1$), $M_{\rm in}^t \equiv M_{\rm in}(M,t)$ is the initial mass that evolves to $M$ at a time $t$. Finally, $dM_{\rm in}/dM$ is the Jacobian that transforms the initial mass to the mass at a time $t$.
It can be readily computed (see Refs.~\cite{Mosbech:2022lfg,Cheek:2022mmy}):
\begin{align}
    \frac{dM_{\rm in}^t}{dM}  =\frac{dM_{\rm in}^t/dt}{dM/dt} = \frac{\varepsilon(M_{\rm in}(M,t)) }{\varepsilon (M)} \frac{M^2}{M_{\rm in}(M,t))^2}\,,
\end{align}
where we have changed the variable from time to scale factor according to, $d/dt \equiv aH d/da$ as $H=\dot{a}/a$.
Since PBHs behave like matter, their energy density is given by the product of the PBH number density and the mass, with the latter evolving in time. Additionally, we include equations for the comoving radiation energy density as well as the Hubble parameter
\begin{equation}\label{eq:Friedmann}
\begin{aligned}
\frac{d \varrho_{\mathrm{R}}}{d a} & = \frac{1}{a\,\Delta} \bigl[4\,(\Delta - 1) \;-\; \Sigma\bigr]
- \frac{n_{\mathrm{BH}}^{\rm in}}{H} \int\frac{\varepsilon_{\rm SM}}{\varepsilon}\frac{dM}{dt} \, f_{\rm PBH} (M_{\rm in}^t)\,\frac{dM_{\rm in}^t}{dM} \,dM\,,\\[6pt]
H^2 & = \frac{1}{3 M_{\mathrm{pl}}^2}\,\bigl(\varrho_{\mathrm{BH}} \,a^{-3} \;+\; \varrho_{\mathrm{R}} \,a^{-4}\bigr)\,,
\end{aligned}
\end{equation}
where $\epsilon_{\rm SM}(M)$ includes only the Standard Model contribution to the evaporation and
\begin{align}
\Sigma = \frac{T}{g_*(T)}\,\frac{dg_*(T)}{dT}\,,
\quad
\Delta = 1 + \frac{T}{3\,g_{*S}(T)}\,\frac{dg_{*S}(T)}{dT}\,,
\end{align}
with $g_*(T)$ and $g_{*S}(T)$ denoting the effective and entropic numbers of relativistic degrees of freedom, respectively.

Since PBH evaporation injects additional entropy into the Universe, the overall entropy is not conserved. Therefore we track the evolution of the ambient temperature $T$ following Refs.~\cite{Lunardini:2019zob,Bernal:2019lpc,Arias:2019uol,Cheek:2022mmy}:
\be\label{eq:TUev}
\frac{dT}{da} \;=\; -\frac{T}{\Delta}\,\biggl\{
\frac{1}{a} - 
\frac{g_{*}(T)}{g_{*S}(T)}\,\frac{d \varrho_{\rm R}/d a}{4\,d \varrho_{\rm BH}/d a}
\biggr\}\,.
\ee
We solve \equasref{eq:Friedmann}{eq:TUev} as a function of the scale factor, typically stopping at $M_{\rm BH}\approx 100 \cdot M_{\rm Pl}$. In practice, the black hole will stop generating a lepton asymmetry before reaching this limit as the effective field theory breaks down beyond a certain scale determined by the condition given in \equaref{eq:breakdown}. We adopt a conservative prescription by assuming symmetric Hawking evaporation from when the effective field description becomes invalid.
This choice is conservative because, in reality, the chemical potential would continue increasing until the black hole fully evaporates, thereby amplifying the evaporation bias toward leptons (or antileptons) and yielding a larger final lepton asymmetry.

The rate of change of the PBH mass and plasma temperature, $T$, will be unchanged from previous calculations \footnote{In principle one should take into account consistently the evolution of the chemical potential as well. However, it will be negligible for most of the PBH evolution time and only become significant at the very end before reaching the breakdown point. 
We have verified numerically that it only induces a negligible correction.
}. 
The evolution of the lepton asymmetry is
\be\label{eq:lepass}
    a H\frac{dn_{\overline{L}-L}}{da} = n_{\mathrm{BH}}^{\rm in} \int\frac{8\pi M_\mathrm{pl}^2}{M}\, \widetilde{\Gamma}_{\rm BH \to \Delta L} \, f_{\rm PBH} (M_{\rm in}^t)\,\frac{dM_{\rm in}^t}{dM} \,d M\,,
\ee
where $\widetilde{\Gamma}_{\rm BH \to \Delta L}$ (see \equaref{gammadef}) is shown in \figref{FigGammaBH}, $n_{\overline{L}-L}$ denotes the leptonic asymmetry number density. As PBH evaporation occurs significantly before the electroweak phase transition, electroweak sphalerons will convert the lepton asymmetry to a baryonic one. 

One can ask if the lepton asymmetry gets washed out or removed via competing processes. If the lepton asymmetry is generated at energies below the scale of the new physics responsible for neutrino masses, then the dominant washout processes are $\Delta L=2$ interactions. Their cross-section can be approximated as
\begin{equation}
\sigma_{\Delta L=2} \,\sim\, \frac{m_\nu^2}{v^4}\,,
\end{equation}
resulting in a rate
\begin{equation}
\Gamma_{\mathrm{washout}} \,\sim\, \frac{m_\nu^2}{v^4}\,T^3\,.
\end{equation}
Comparing this rate to the Hubble expansion parameter, one finds that the washout processes fail to reach equilibrium whenever
\begin{equation}
\Gamma_{\mathrm{WO}} 
\,=\, \frac{m_\nu^2}{v^4} \, T^3
\,\lesssim\,
0.33\,\sqrt{g_{\star}}\,\frac{T^2}{M_\mathrm{pl}}
\;\;\Longrightarrow\;\;
T\,\gtrsim\, 5\times 10^{11}\,\text{GeV},
\end{equation}
assuming $m_\nu \simeq 0.1\,\text{eV}$. Hence, as long as the Universe’s temperature remains below this temperature during the bulk of the lepton-asymmetry production, the washout never becomes efficient, and it can safely be ignored in the Boltzmann equation of \equaref{eq:lepass}.

\section{Results}\label{sec:results}
\begin{figure}
    \centering
    \includegraphics[width=1\textwidth]{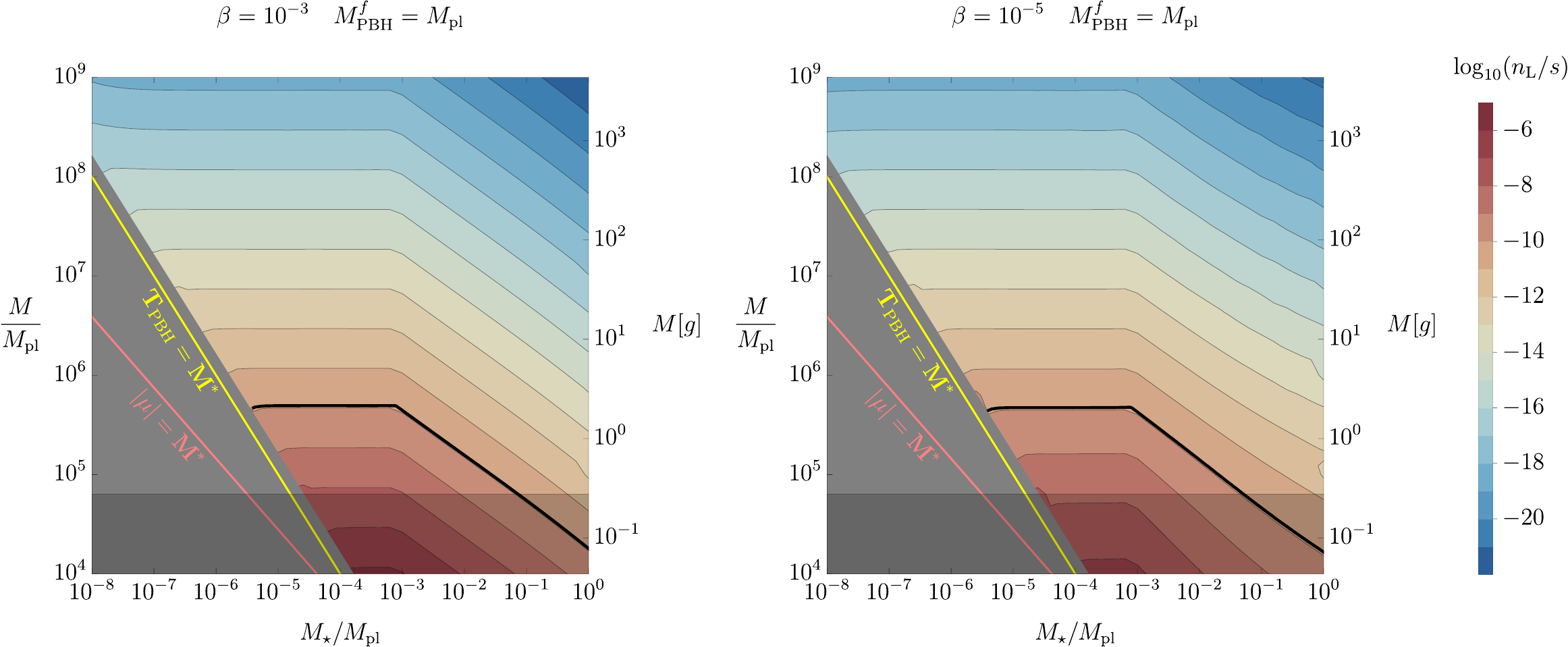}
    \caption{Lepton asymmetry yield $n_L/s$ for $\beta=10^{-3}$ (left) and $\beta=10^{-5}$ (right) assuming a monochromatic mass spectrum obtained by solving the full Boltzmann equations. We numerically evolve the equation down to a final PBH mass equal to $M_{\rm pl}$. See \figref{fig1} for a description of the figure.}
    \label{fig7}
\end{figure}

To obtain the total lepton asymmetry, we solve the complete set of \equasref{eq:Friedmann}{eq:lepass} accounting for the full evolution of the Universe with the exact greybody factors for different PBH mass distributions. Let us discuss the monochromatic scenario first. In \figref{fig7}, we scan over the parameter space given by the cutoff scale $M_\star$ and the PBH mass at formation. We show the results for a value of the initial abundance $\beta=10^{-3}$ (left) and $\beta=10^{-5}$ (right) where the black line indicates the central value of the baryon asymmetry. We observe the same qualitative features as in the results in \figref{fig2} with the analytic treatment. Namely, a constant leptonic yield for a fixed initial PBH mass and cutoff $M_{\star}<10^{-3}M_{\rm pl}$, and a decrease at larger values of the cutoff scale. Also, larger initial PBH masses lead to a lower final leptonic yield and vice versa. Indeed, for a fixed abundance, a population of heavier PBHs corresponds to a lower number density, hence there will be less black holes contributing to the lepton asymmetry. Also, there will be a larger entropy dilution, as heavier PBHs will convert a larger fraction of their mass into radiation not contributing to the asymmetry. There is however an overall noticeable difference. We observe a very significant increase of the leptonic yield below $M \sim 10^{6.8} M_{\rm pl}$ ($M \sim 10^{8.8} M_{\rm pl}$) for $\beta=10^{-3}$ ($\beta=10^{-5}$), and up to a factor $\sim4$ ($\gtrsim10$) for the lowest PBH initial masses. We have checked that for small initial PBH masses the entropy dilution becomes irrelevant and the difference is due to the use of the exact greybody factors. On the other hand, the full numerical analysis results in a gradual reduction of the leptonic yield as we move to larger PBH initial masses, and up to 70$\%$ less compared to the analytical values. For these values of the PBH mass, even if the enhancement from the exact greybody factor still applies, the entropy dilution becomes dominant. The difference is then due to the different prescriptions used to compute the entropy dilution, namely \equaref{eq:entropydil} in the analytical treatment versus the full numerical setup. 

\begin{figure}
    \centering
    \includegraphics[width=1\textwidth]{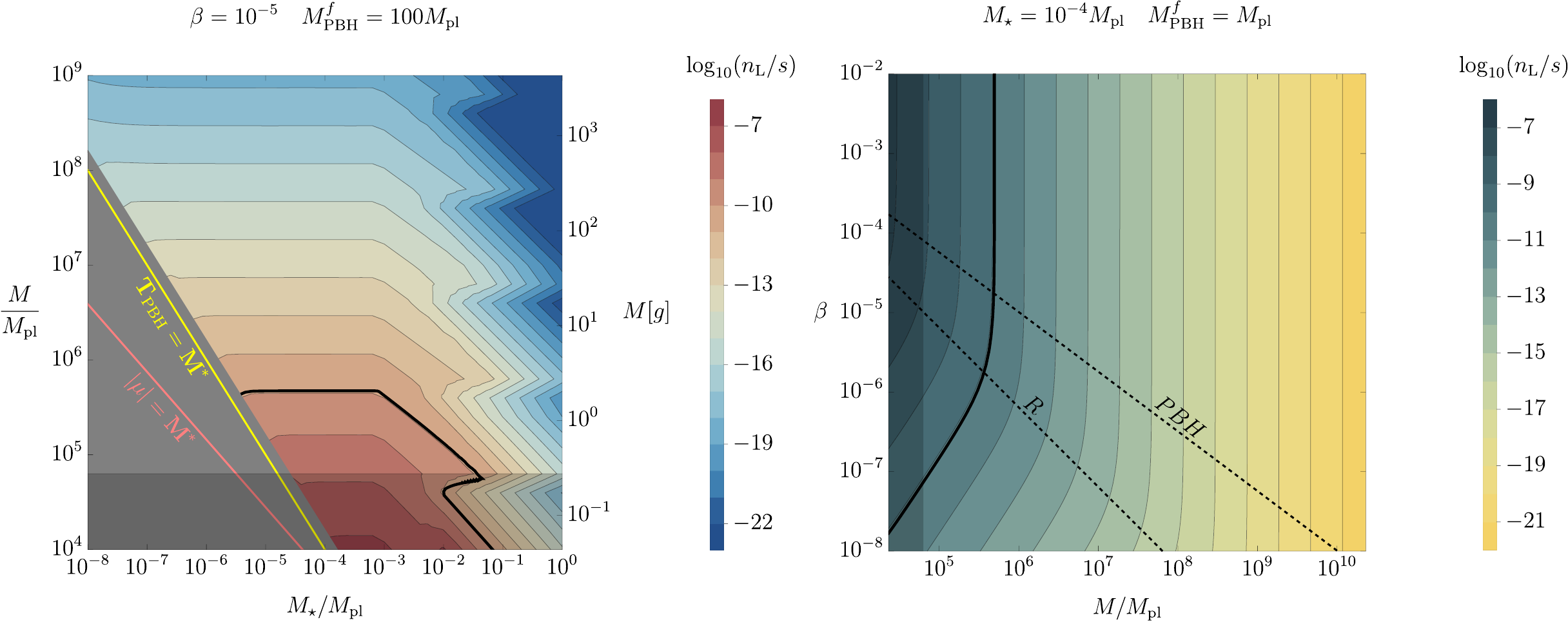}
    \caption{Lepton asymmetry yield $n_L/s$ for $\beta=10^{-5}$ and stopping the mass evolution at $M^f_{\rm PBH}=100M_P$ (left) and scanning over $\beta$ for a fixed value of the cutoff scale $M_{\star}=10^{-4}M_P$ (right). The dashed line indicates the line satisfying $1=\beta T_0/T_{\rm ev}$, assuming radiation domination (R) or PBH domination (PBH). Below this line entropy dilution is negligible. See \figref{fig1} for a description of the figure.}
    \label{fig8}
\end{figure}

In \figref{fig7} we evolve the PBH masses down to the Planck mass, after which Hawking evaporation is assumed to halt. We show the results when changing this prescription and evolving down to a final mass $M^f_{\rm PBH} = 100M_{\rm pl}$ for $\beta=10^{-5}$ in \figref{fig8} left. 
Since we impose a cutoff at a fixed final mass $M^f$, the numerical evolution shows that, for some initial PBH masses, the evolution stops before reaching $M_{\rm min}$, while for others the cutoff is reached first. Our algorithm iteratively solves the PBH system in steps, reducing the mass from its initial value to 10\% of that value. As a result, the point at which either $M_{\rm min}$ or $M^f_{\rm PBH}$ is reached varies periodically, producing the zigzag pattern observed in this figure. Therefore, apart from this numerical artifact, no significant differences are seen between \figref{fig7} and \figref{fig8}.

We also scan over several values of $\beta$ for a given choice of the cutoff scale $M_{\star}=10^{-4}M_{\rm pl}$ in \figref{fig8} right. Interestingly, the resulting leptonic yield becomes independent of the PBH abundance $\beta$ at large initial PBH masses as well as for $\beta \gtrsim 10^{-5}$ (above the dashed line in \figref{fig8} right). In this region $1\ll\beta T_0/T_{\rm ev}$, and the universe goes through an epoch of early PBH domination before PBH decay. Therefore, from the analytical treatment, in particular \equaref{eq:lepasymdil} and \equaref{eq:tev_pbh}, one can see that for a fixed initial PBH mass the resulting leptonic yield indeed becomes independent since $T_{\rm ev}\propto\beta^{-1/3}$. On the other hand, when $1\gg\beta T_0/T_{\rm ev}$, the entropy dilution is negligible and the leptonic yield increases linearly with $\beta$.

\begin{figure}[t!]
    \centering    \includegraphics[width=\textwidth]{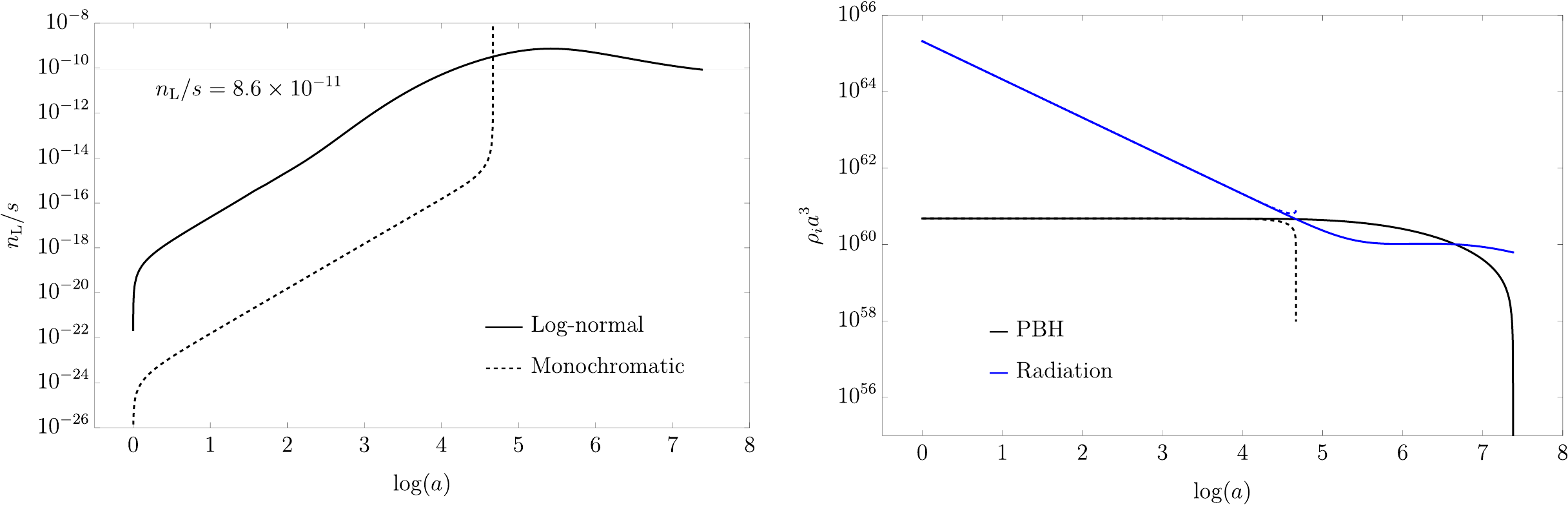}
    \caption{Results for a log-normal mass distribution with $M_{\rm scl}=0.235$g and $\sigma=1$, and parameters $\beta=10^{-5}$ and $M_{\star}=10^{-3} M_{\rm pl}$. The dashed lines correspond to a monochromatic mass function with the same value of $M_{\rm scl}$. We show the leptonic yield as a function of the scale factor on the left panel, and the radiation (blue) and PBH (black) energy density evolution on the right.}
    \label{fig5}
\end{figure}



\begin{figure}[t!]
    \centering
\includegraphics[width=\textwidth]{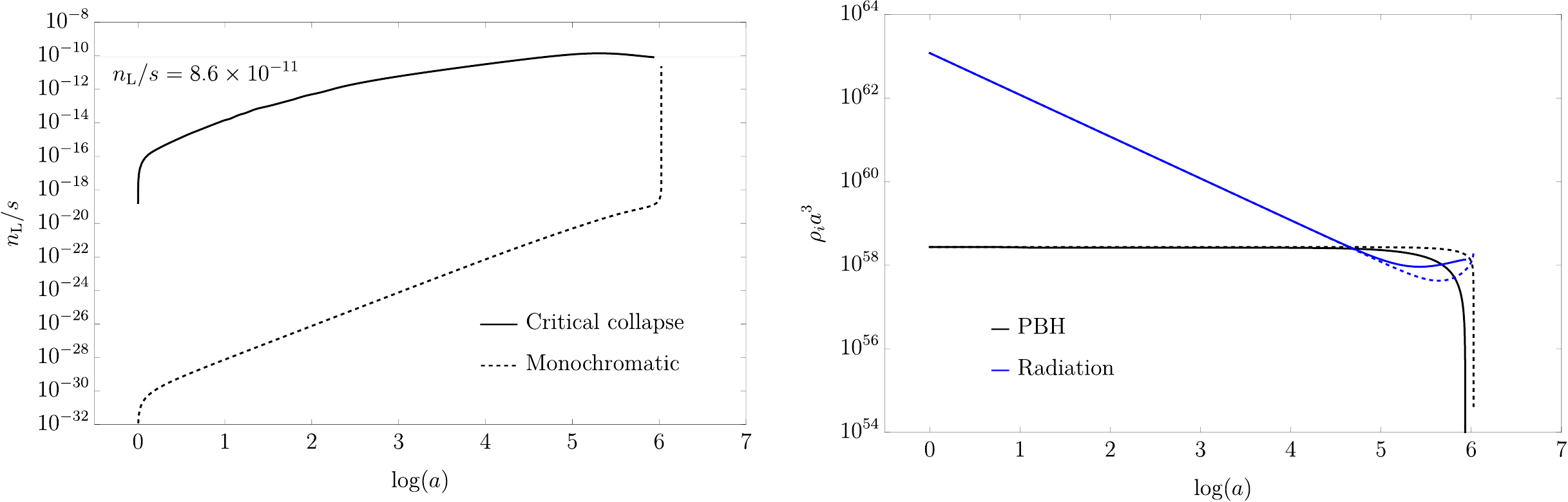}
    \caption{Results for a generalised critical collapse mass distribution with $M_{\rm scl}=3.125$g, $\kappa=2.001$ and $\eta=1000$, and parameters $\beta=10^{-5}$ and $M_{\star}=10^{-3} M_{\rm pl}$. The dashed lines correspond to a monochromatic mass function with the same value of $M_{\rm scl}$. We show the leptonic yield as a function of the scale factor on the left panel, and the radiation (blue) and PBH (black) energy density evolution on the right.}
    \label{fig6}
\end{figure}

Moving on to extended mass distributions, in \figsref{fig5}{fig6}, we show the results for two example scenarios with a log-normal and a generalised critical collapse mass distribution, respectively, and compare them with the monochromatic case (dashed line). The right panel shows the energy density of radiation and PBH in blue and black, respectively. In the left panel, we show the leptonic yield $n_{\text{L}}/s$.
In both cases the Universe presents an early phase of PBH domination starting around $\log(a)\sim 5$, as observed in the right panel of \figref{fig6}. Note that this phase is also present for a monochromatic mass function for large values of $M_{\rm scl}$. The duration of this epoch is shorter for the critical collapse case, due to the less extended high-mass tail compared to the log-normal mass function, which corresponds to heavier PBHs that evaporate later. On the other hand, the enhanced low-mass tail leads to a larger leptonic yield already at very early times compared to the log-normal scenario. In both cases, the leptonic yield increases initially at a similar pace as in the monochromatic case, essentially given by the linear increase of the lepton asymmetry rate discussed in \figref{FigGammaBH}.
As the lighter PBHs complete their evaporation, the leptonic yield gradually stabilises at the observed value $n_L/s\sim10^{-10}$. The resulting curve is smoother compared to the monochromatic case as PBHs completely evaporate at different moments over some extended period of time rather than having a population of PBHs evaporating at the same instant, where the dashed line suddenly increases by several orders of magnitude. Note that there is a slight decrease in the very end due to the effect of heavier PBHs producing a large amount of radiation-like particles that do not contribute to the lepton asymmetry. We also observe the same qualitative features with a power-law mass distribution, for which the observed baryon asymmetry can also be reproduced.

\section{Conclusion}\label{sec:conclusions}

In this paper we have revisited the role of light PBHs in the context of baryogenesis (and leptogenesis). In particular, we studied the evaporation of a population of PBHs in the framework of an effective field theory with a dimension-eight, CP-violating operator coupling the derivative of the Kretschmann scalar with a baryon (or lepton) number-violating current, see \eqref{eq:CPVop}. The evolution of the PBH mass generates a non-zero time derivative that induces an effective chemical potential $\mu$ on the PBH horizon, biasing Hawking evaporation towards more baryons (or leptons) over antibaryons (or antileptons).

We have improved upon previous results on several fronts. Firstly, we performed our computations using 
$\mu$-dependent greybody factors. As shown in \figref{FigGammaBH}, the resulting lepton asymmetry rate of a single PBH with exact greybody factors can differ substantially from results obtained by approximating the greybody factors in the high- or low-energy limits. Secondly, we considered the full PBH evaporation spectrum and, in particular, the contribution from evaporation into photons, which had previously been neglected. We find this component to play a crucial role through entropy dilution, which constrains the parameter space capable of reproducing the observed baryon asymmetry, as illustrated in \figref{fig8} (right). Finally, we consistently solved the complete set of equations describing the full evolution of the Universe, including radiation and a PBH population characterised by an extended mass distribution. We presented two examples in \figref{fig5} and \figref{fig6}, corresponding to a critical collapse and a log-normal mass function, respectively. In both cases, the observed baryon asymmetry is successfully reproduced.

In conclusion, our findings demonstrate that the observed baryon asymmetry can be explained by a biased evaporating population of light PBHs. Our results hold for the three types of mass distributions we considered: critical collapse, log-normal, and power-law. Moreover, based on our study of the monochromatic mass function, we expect our conclusions to remain valid across a wide range of initial PBH abundances $\beta$ and effective field theory cutoffs $M_{\star}$, with entropy dilution representing the main physical limitation. At the same time, as discussed in Ref.~\cite{Hamada:2016jnq}, our results should be regarded as conservative, since PBHs are expected to continue generating baryon asymmetry beyond the scale $M_{\star}$, where our effective description ceases to be applicable. A complete UV theory could therefore open additional regions of parameter space capable of accounting for the observed baryon asymmetry. Further work is needed to explore the possible origin of the CP-violating higher-dimensional operators.

\section*{Acknowledgments}
 JT would like to thank the Quantum Field Theory Centre at the University of Southern Denmark for their hospitality during the completion of this work.
 YFPG has been supported by the Consolidaci\'on Investigadora grant CNS2023-144536 from the Spanish Ministerio de Ciencia e Innovaci\'on (MCIN) and by the Spanish Research Agency (Agencia Estatal de Investigaci\'on) through the grant IFT Centro de Excelencia Severo Ochoa No CEX2020-001007-S.

\noindent

\appendix






\section{Greybody Factors for fermions including chemical potential}
\label{sec:AppGBfactors}

Let us consider the Dirac equation in the Schwarzschild background, including a new interaction in the form of a chemical potential. The action, following Ref.~\cite{Hamada:2016jnq} is
\begin{align}
    S = \int \sqrt{-g}\, d^4x \overline{\psi}\left(i\partial_\mu - \frac{c}{M_{\star}^4}\partial_\mu{\cal \mathcal{K}}\right)\gamma^\mu \psi\,,
\end{align}
where ${\cal \mathcal{K}} = R_{\alpha\beta\gamma\delta}R^{\alpha\beta\gamma\delta}$ is the Kretschmann scalar for the Schwarzschild metric, given by
\begin{align}
    \mathcal{K} = \frac{3 M^2}{4 \pi^2 M_{\mathrm{pl}}^4 r^6},
\end{align}
where $M$ is the instantaneous BH mass. The Dirac equation in the two-component spinor notation is given by~\cite{Page:1976jj}
\begin{align}
    i(\nabla_{B B^\prime} + i A_{B B^\prime})P^B - \mu_e Q_{B^\prime}^* &= 0\,,\\
    i(\nabla_{B B^\prime} - i A_{B B^\prime})Q^B - \mu_e P_{B^\prime}^* &= 0\,,
\end{align}
where $\nabla_{B B^\prime}$ is the covariant differentiation in te pseudo-Riemannian geometry related to the Schwarzschild spacetime, $A_{B B^\prime}$ is the vector field associated to the chemical potential, $\mu_e$ is the fermion mass, $P^B, Q^B$ are the two component spinors, such that
\begin{align*}
    \psi = \begin{pmatrix}
                P^B\\
                Q_{B^\prime}
            \end{pmatrix}\,.
\end{align*}
We use the Newman-Penrose formalism with a null tetrad ($\mathbf{l, n, m, M_{\star}}$), we get the system of equations,  see Ref.~\cite{Page:1976jj}
\begin{subequations}
    \begin{alignat}{3}
        (D - \epsilon - \rho + i \mathbf{A}\cdot\mathbf{l}) P^0 &+ (\delta^*+\pi-\alpha+i\mathbf{A}\cdot\mathbf{M_{\star}})P^1 &&= &&i \mu_e Q^{*1^{\prime}},\\
        (\bigtriangleup + \mu - \gamma + i \mathbf{A}\cdot\mathbf{n}) P^1 &+ (\delta+\beta-\tau+i\mathbf{A}\cdot\mathbf{m})P^0 &&= -&&i \mu_e Q^{*0^{\prime}},\\
        (D - \epsilon^* - \rho^* + i \mathbf{A}\cdot\mathbf{l}) Q^{*0^{\prime}} &+ (\delta+\pi^*-\alpha^*+i\mathbf{A}\cdot\mathbf{m})Q^{*1^{\prime}} &&= -&&i \mu_e P^{1},\\
        (\bigtriangleup + \mu^* - \gamma^* + i \mathbf{A}\cdot\mathbf{n})Q^{*1^{\prime}} &+ (\delta^*+\beta^*-\tau^*+i\mathbf{A}\cdot\mathbf{m})Q^{*0^{\prime}} &&= &&i \mu_e P^0.
    \end{alignat}
\end{subequations}
For the Schwarzschild spacetime, one has the contravariant four vectors
\begin{align}
    \mathbf{l}^\mu = \frac{1}{\Delta}(r^2, \Delta, 0, 0),\quad\mathbf{n}^\mu = \frac{1}{2r^2}(r^2, -\Delta, 0,0),\quad \mathbf{m}^\mu =\frac{1}{\sqrt{2}r} (0,0,1, i\csc\theta)
\end{align}
where $\Delta = r(r-M/(4\pi M_{\mathrm{pl}}^2))$.
The components relevant for the Dirac equation are
\begin{align}
    D = \mathbf{l}^\mu \partial_\mu = \frac{r^2\partial_t}{\Delta} + \partial_r,\quad \bigtriangleup= \mathbf{n}^\mu \partial_\mu=-\frac{\Delta}{2r^2}D^*,\quad \delta = \mathbf{m}^\mu \partial_\mu= \frac{1}{\sqrt{2}r}(\partial_\theta +i \csc\theta\partial_\phi) = \delta^*,\\
    \epsilon = \pi = \tau = 0, \rho = -\frac{1}{r}, \beta = -\alpha =\frac{1}{2\sqrt{2}}\frac{\cot\theta}{r}, \mu =-\frac{1}{2} +\frac{M}{8\pi M_{\mathrm{pl}}^2r^2}, \gamma = \frac{M}{8\pi M_{\mathrm{pl}}^2r^2},
\end{align}
with $\omega$ represents the energy. Taking $A_\mu = -(A_0, 0, 0,0)$, $A_0$ is the time component of the chemical potential four vector, we have
\begin{align}
    \mathbf{A}\cdot\mathbf{l} = -\frac{A_0 r^2}{\Delta},\quad \mathbf{A}\cdot\mathbf{n} = -\frac{A_0}{2},\quad \mathbf{A}\cdot\mathbf{m}=\mathbf{A}\cdot\mathbf{M_{\star}}=0.
\end{align}
Separating $r$ and $\theta$ variables by defining
\begin{subequations}
    \begin{align}
        P^0 &= \frac{1}{r}e^{i(\omega t + m\phi) R_{-1/2}(r) S_{-1/2}(\theta)},\\
        P^1 &= e^{i(\omega t + m\phi) R_{+1/2}(r) S_{+1/2}(\theta)},\\
        Q^{*1^\prime} &= e^{i(\omega t + m\phi) R_{+1/2}(r) S_{-1/2}(\theta)},\\
        Q^{*0^\prime} &= -\frac{1}{r}e^{i(\omega t + m\phi) R_{-1/2}(r) S_{+1/2}(\theta)}.
    \end{align}
\end{subequations}
Here $R_{\pm 1/2}(r)$ and $S_{\pm 1/2}(\theta)$ are functions depending only on the radial and angular variables. 
Since the new interaction assumed here is spherically symmetric, we have that the angular functions $S_{\pm 1/2}(\theta)$ are the same as in the standard Schwarzschild case, see e.g.~Ref.~\cite{Doran:2005vm,Dolan:2006vj}.
To compute the greybody factors, we need to solve the equations for $R_{\pm 1/2}(r)$. Letting $R_{-1/2}\equiv R_{1}$ and $R_{+1/2} = (2/\Delta)^{1/2}R_2$
\begin{subequations}
    \begin{align}
        \left(\partial_r + \frac{i K }{\Delta}\right)R_1 &= (\lambda + i\mu_e r)\Delta^{-1/2} R_2,\\
        \left(\partial_r - \frac{i K }{\Delta}\right)R_1 &= (\lambda - i\mu_e r)\Delta^{-1/2} R_1,
    \end{align}
\end{subequations}
where $K = (\omega - A_0)r^2$, and $\lambda = j+1/2$, $j\in [1/2, 3/2, 5/2, \ldots]$, the angular eigenvalue. Performing the same transformation as in Ref.~\cite{Page:1977um}, $R_1 = G+iF$, $R_2=G-iF$, we get
\begin{subequations}
    \begin{align}
        (\Delta^{1/2}\partial_r - \lambda)G + (-K \Delta^{-1/2} + \mu_e r)F = 0,\\
        (\Delta^{1/2}\partial_r + \lambda)F + (+K \Delta^{-1/2} + \mu_e r)G = 0.
    \end{align}
\end{subequations}
Since the shape of these equations is the same as the ones given in Ref.~\cite{Page:1977um}, to obtain the greybody factors, we follow the procedure detailed in the Appendix of Ref.~\cite{Page:1977um}, with the appropriate changes due to the additional potential $A_0$, given by
\begin{align}
    A_0  &= \frac{c}{M_{\star}^4}\partial_t \mathcal{K}\notag\\
    &= \frac{3 c}{2 \pi^2 M_{\star}^4} \frac{1}{r^6 M_{\mathrm{pl}}^4}M\frac{dM}{dt}\notag\\
    &= -\alpha \left(\frac{r_S}{r}\right)^6,
\end{align}
where
\begin{align}
    \alpha = \frac{3 c}{2} (8\pi)^6 \varepsilon(M) M_\mathrm{pl} \left(\frac{M_\mathrm{pl}}{M}\right)^7 \left(\frac{M_\mathrm{pl}}{M_{\star}}\right)^4 = \left.\mu\right|_{r=r_S}
\end{align}
being $r_S = M/(4 \pi M_\mathrm{pl}^2)$ the Schwarzschild radius and $\mu \left.\mu\right|_{r=r_S}$ is the chemical potential evaluated at the horizon, as given in Eq. (21) of Ref.~\cite{Hamada:2016jnq}. 
When performing the variable transformation given in Ref.~\cite{Page:1977um}, we find that we need to work with the dimensionless variable $\widetilde{\alpha}$,
\begin{align}
   \widetilde{\alpha} = 3 c (8\pi)^5 \varepsilon(M) M_\mathrm{pl} \left(\frac{M_\mathrm{pl}}{M}\right)^6 \left(\frac{M_\mathrm{pl}}{M_{\star}}\right)^4 = \left(\frac{M_{\rm cr}}{M}\right)^6,
\label{Mcriticaldef}
\end{align}
which is proportional to $\left.\mu\right|_{r=r_S}/T_{\rm BH}$ given in the same paper, apart from a factor of $4\pi$.

\begin{figure}
    \centering
    \includegraphics[width=0.8\textwidth]{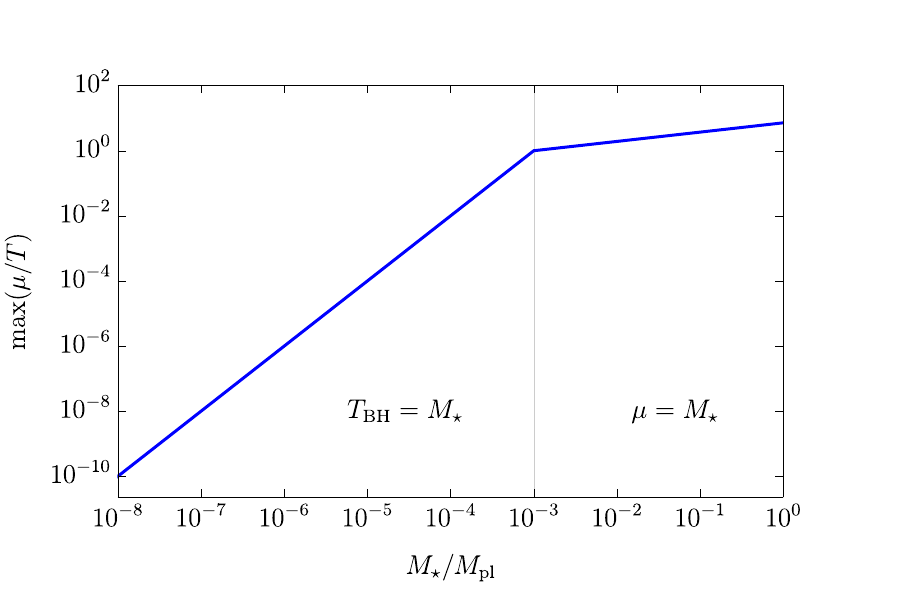}
    \caption{Maximum value for $\left.\mu\right|_{r=r_S}/T_{\rm BH}$ as a function of the cut-off scale $M_{\star}$ ($M_{\rm P}$ is the reduced Planck mass). On the left (right) of the horizontal grey line, the maximum value is given by the condition $T_{\rm BH}=M_{\star}$ ($\left.\mu\right|_{r=r_S}=M_{\star}$).}
    \label{fig0} 
\end{figure}

The maximum value for $\left.\mu\right|_{r=r_S}/T_{\rm BH}$ can in principle go way beyond 1 as the black hole evaporates and its mass falls below $M_{\rm cr}$. An absolute upper bound can be derived by assuming the maximum value for the cutoff scale $M_{\star}$ and the minimum value for the black hole mass $M_{\rm BH}$ for which our current physical theories still apply. This corresponds to $M_{\star}=M_{\rm p}$ and $M_{\rm BH}=M_{\rm p}$, which leads to $\left.\mu\right|_{r=r_S}/T_{\rm BH} \sim 10^6$, where we used Eq.21 and 22 given in Ref.~\cite{Hamada:2016jnq}. However, this value will not be reached because, as the black hole evaporates, the temperature and the chemical potential increase until $T_{\rm BH}=M_{\star}$ and/or $\left.\mu\right|_{r=r_S}=M_{\star}$ is satisfied, beyond which the effective theory is no longer valid. For a given value of the cut-off $M_{\star}$, we can only track the black hole evolution down to a mass given by Eq. 28 in Ref.~\cite{Hamada:2016jnq}. This sets the maximum value for $\left.\mu\right|_{r=r_S}/T_{\rm BH}$ for that particular value of $M_{\star}$. In Figure~\ref{fig0} we show the maximum values in the range of interest for $M_{\star}$.

\bibliographystyle{JHEP}
\bibliography{Bibliography.bib}

\end{document}